\documentclass[12pt]{article}
\usepackage{amsmath}
\usepackage{times}
\usepackage{graphicx}
\usepackage{subcaption}
\graphicspath{{./Figs/}}
\usepackage{color}
\usepackage{multirow}
\usepackage[authoryear,round]{natbib}
\usepackage{rotating}
\usepackage{bbm}
\usepackage{latexsym}
\usepackage{setspace}

\usepackage[capitalise]{cleveref}

\newcommand{\captionfonts}{\normalsize}

\makeatletter  
\long\def\@makecaption#1#2{%
  \vskip\abovecaptionskip
  \sbox\@tempboxa{{\captionfonts #1: #2}}%
  \ifdim \wd\@tempboxa >\hsize
    {\captionfonts #1: #2\par}
  \else
    \hbox to\hsize{\hfil\box\@tempboxa\hfil}%
  \fi
  \vskip\belowcaptionskip}
\makeatother   

\begin{document}
\hspace{13.9cm}1

\ \vspace{20mm}\\

{\LARGE Reliability of Event Timing in Silicon Neurons}

\ \\
{\bf \large Tai Miyazaki Kirby$^{\displaystyle 1}$, Luka Ribar$^{\displaystyle 1}$, Rodolphe Sepulchre$^{\displaystyle 1}$}\\
{$^{\displaystyle 1}$Engineering Department, University of Cambridge.}\\
%


\thispagestyle{empty}
\markboth{}{NC instructions}
\ \vspace{-0mm}\\
%
\begin{center} {\bf Abstract} \end{center}
Analog, low-voltage electronics show great promise in producing silicon neurons (SiNs) with unprecedented levels of energy efficiency. Yet, their inherently high susceptibility to process, voltage and temperature (PVT) variations, and noise has long been recognised as a major bottleneck in developing effective neuromorphic solutions. Inspired by spike transmission studies in biophysical, neocortical neurons, we demonstrate that the inherent noise and variability can coexist with reliable spike transmission in analog SiNs, similarly to biological neurons. We illustrate this property on a recent neuromorphic model of a bursting neuron by showcasing three different relevant types of reliable event transmission: single spike transmission, burst transmission, and the on-off control of a half-centre oscillator (HCO) network.

\section{Introduction}

Since the seminal work of Carver Mead \citep{mead1989}, the field of neuromorphic engineering has leveraged biological inspiration to deliver great advances in the fields of novel sensing, actuation, and processing devices \citep{mead1990,indiveri2011,furber2016,boahen2017,gallego2020}. Throughout its history, one of the biggest focuses in neuromorphic engineering has been the development of silicon neurons (SiNs) - silicon hardware that is designed to mimic the behaviour of single, biophysical neurons.  SiNs can be interconnected via silicon synapse circuits to create more complex neural networks.  Traditionally, SiNs have been implemented with analog circuits primarily based upon the sub-threshold operation of MOS transistors \citep{mead1989,mahowald1991,vanschaik2001,simoni2004,hynna2007,yu2010}. However, it has long been recognised that one of the biggest problems impeding the realisation of robust and scalable SiN networks is the susceptibility of analog circuit implementations to process, voltage and temperature (PVT) variations and noise (e.g. transistor mismatch and thermal noise). Many techniques have been developed to compensate for these PVT variations and noise, some of them being:

\begin{itemize}
	\item Algorithm-based tuning \citep{green2017, shuo2011}
	\item Adaption/learning \citep{cauwenberghs1999}
	\item Data assimilation \citep{wang2017}
	\item Specific VLSI layout techniques \citep{liu2002} and VLSI circuits manufactured to high-precision.
\end{itemize}

Alternatively, other approaches have focused on producing SiNs with either primarily digital or mixed analog-digital operation, for the higher precision and reproducibility offered by digital circuits \citep{cassidy2008,serrano-gotarredona2009,kim2015,wijekoon2008}. Such implementations, however, are orders of magnitude less energy-efficient compared to their entirely analog counterparts for two reasons.  Firstly, digital circuits are built from transistors operated in the saturation regime.  Secondly, digital implementations cannot take advantage of the similarities between transistor physics and the biophysics underlying neuronal dynamics.  Instead, neuronal dynamics must be simulated numerically - an additional departure from the biophysical inspiration.

This letter, however, argues that PVT (particularly temperature) variability and noise in analog SiNs is not necessarily problematic. In fact, similarly to their biological counterparts, the variability/noise of analog SiNs does not preclude the design of reliable circuits. In biology, noise and variability are pervasive in the analog signals transmitted from one neuron to another, yet inter-neuron communication at the network scale is reliable (i.e. event timings are robust). This robustness of event timings in networks owes to the threshold property of excitable neurons. We demonstrate the same property in SiNs, by reproducing the famous reliability experiments by Mainen and Sejnowski, originally performed in vitro on neocortical neurons \citep{mainen1995}.

In these experiments, a rat neocortical neuron was repeatedly injected with the same current waveform across multiple trials and the resulting spike trains were recorded in a current-clamp configuration. When the neuron was injected with a DC current step, spike timings would become rapidly unreliable over the course of the stimulus, demonstrating high variability between trials.  In contrast, when presented with the same rapidly fluctuating stimulus (generated using a Gaussian white noise process) in each trial, spike timings were highly consistent across trials and over the course of the stimulus. In other words, the spike transmission was highly \textit{reliable} when the neuron received input currents similar to those it would typically receive in vivo.  Reliability, here, refers to the robust alignment of spike timings across multiple trials, in response to the same fluctuating input presented in each trial.  These results substantiated earlier observations of reliability in R2 neurons of the Aplysia californica - neurons capable of both spiking and bursting \citep{bryant1976}. If analog SiNs can capture this property, the experiments suggest that a reliable, event-based neuromorphic circuit can be designed with noisy and imperfect low-power, analog primitives.

This letter presents two contributions. Firstly, we reproduce the reliability experiment in SiNs, showing that reliability can coexist with variability in analog spiking circuits exactly like in biological neurons. As a second contribution, we show that reliability is not confined to spiking but extends to the event-based response of any excitable system. To that end, we replicate the reliability experiment on a bursting neuron and on an elementary two-neuron rhythmic circuit. In each situation, we adapt the temporal scale of the input fluctuations to match those of the excitable events, reproducing the reliability experiment at various timescales.

The letter is organized as follows. We start by outlining the architecture of the analog neuromorphic circuit used to study reliability, as well as the experimental set-up used to reproduce the Mainen and Sejnowski experiments for three different temporal scales: spiking in an individual SiN, bursting in an individual SiN, and the anti-phase oscillatory bursting of the half-centre oscillator (HCO), a fundamental pattern-generating network. We present and discuss the results of the three different experiments, and their importance for the design and analysis of neuromorphic hardware.

\section{Methodology}

\subsection{Neuromorphic Model}

In the experiments, we used the recently introduced bursting neuromorphic circuit developed by \citep{ribar2019}. The model represents a simplified neuronal membrane that has the essential neuromodulation mechanisms that enable robust neuromodulation between bursting and spiking regimes. The model is easily reproducible in analog hardware, thus it offers a simple way to test the reliability of spike transmission.

The model is based on the classical Hodgkin-Huxley model and consists of a parallel interconnection of passive membrane elements (a capacitor and a resistor), with simplified models of individual ionic currents, each having the following form:
\begin{equation}
	\label{eq:feedback_current}
	\begin{aligned}
		I_{x}^{\pm} &= \pm \alpha \tanh(V_x - \delta), \\
		\tau_x \dot{V_x} &= V - V_x.
	\end{aligned}
\end{equation}
The output current $I_x^{\pm}$, is defined by its timescale (indicated by the subscript $x$) and its sign (indicated by the superscript). The timescale of the current is characterised by $\tau_x$ and the corresponding first-order filter, which produces a lagged version of the membrane voltage $V_x$. The sign for the current, on the other hand, indicates whether the current has a monotonically increasing or decreasing current-voltage relationship, depending on whether the sign is positive or negative respectively. The parameter $\alpha > 0$ is analogous to the maximal conductance, controlling the gain of the output current. Finally, the parameter $\delta$ controls the voltage range in which the element is \textit{active} (not saturated).

This simplified model of an ionic current is convenient as it captures the two fundamental properties of biophysical ionic currents: they act in well-defined timescales and specific voltage ranges. Thus, the effect of different ionic currents can be specified solely by setting \textit{how fast} it acts ($\tau_x$), \textit{where} in the voltage range it acts ($\delta$), and whether the current tends to \textit{increase} the membrane voltage (sign -), or \textit{decrease} it (sign +).

\begin{figure}[h]
	\centering
	\includegraphics[width=1.0\linewidth]{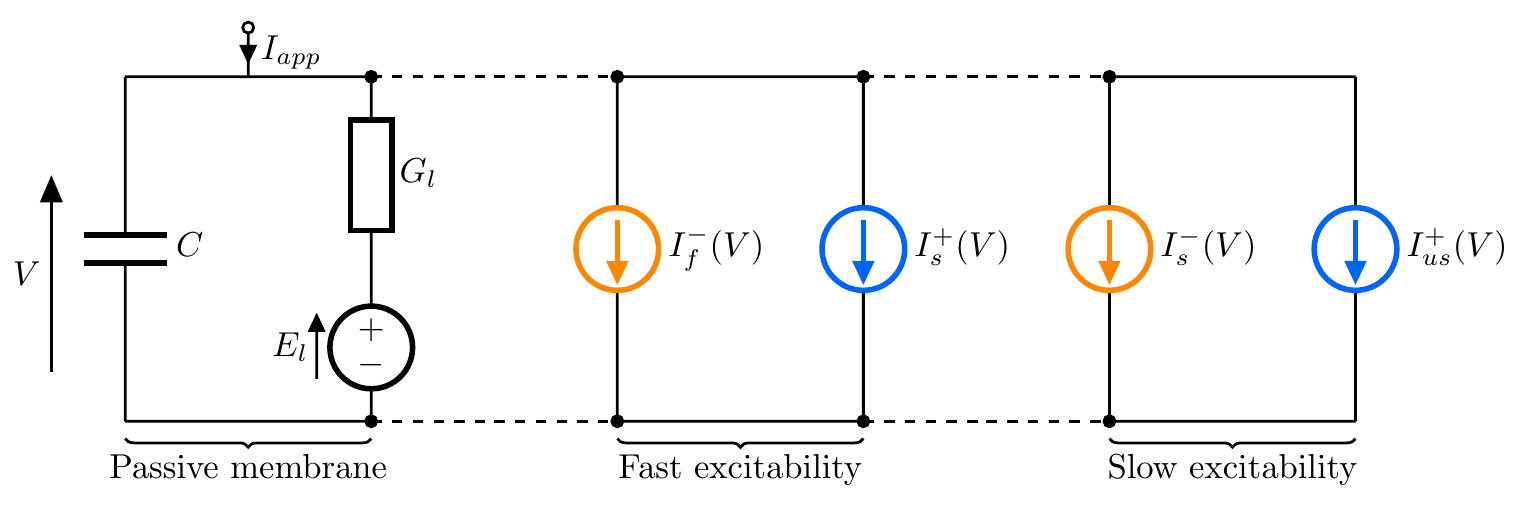}
	\caption{\doublespacing A circuit diagram of the bursting model, consisting of a passive membrane capacitance and conductance, and four ionic currents: $I_f^-$, $I_s^+$, $I_s^-$ and $I_{us}^+$.  Diagram adapted from \citep{ribar2021}.}
	\label{bursting model}
\end{figure}

The bursting model (\cref{bursting model}) consists of \textit{four} currents, capturing the effects of sodium and potassium currents, as well as the slower currents such as calcium and calcium-dependent potassium currents that appear in biophysical bursting models \citep{franci2017}. The model's membrane equation is given by:
\begin{equation}
	\label{eq:membrane}
	C \dot{V} = - (V - E_l) G_l - I_f^-(V) - I_s^+(V) - I_s^-(V) - I_{us}^+(V) + I_{app},
\end{equation}
where $V$ is the membrane voltage, $C$ is the membrane capacitance, $G_l$ and $E_l$ are the conductance of the passive leak current and its reversal potential, $I_f^-$, $I_s^+$, $I_s^-$ and $I_{us}^+$ are the four ionic currents, and $I_{app}$ is the externally applied current. The currents act in three distinct timescales: fast ($f$), slow ($s$) and ultra-slow ($us$), capturing the fast timescale effects required for spike generation (fast and slow nature of sodium and potassium currents, respectively), and the slower timescale effects that produce the bursting patterns (ultra-slow nature of calcium-activated potassium currents). Thus, we have that $\tau_f \ll \tau_s \ll \tau_{us}$.  These time constants are typically fixed during operation.

The behaviour of the circuit is mainly controlled by modulating the maximal conductance parameters of the individual ionic currents ($\alpha$). In particular, we use the gain of the \textit{slow negative conductance} ($I_s^-$) as the main control parameter that sets the behaviour of the neuron as either spiking or bursting, as described in \citep{ribar2019}.

\subsubsection{Half-Centre Oscillator} \label{section: HCO}

To investigate whether the concept of reliability can be scaled up to elementary oscillatory networks, we use the previously described neuromorphic model to create an HCO. The network consists of a mutual interconnection of two neurons with inhibitory synapses, so that the neurons oscillate in anti-phase \citep{bucher2015,ribar2021}. For synaptic elements we use a similar model as the one previously described for the ionic currents \cref{eq:feedback_current}, so that each synapse is modelled as:
\begin{equation}
	\label{eq:synapse}
	\begin{aligned}
		I_{syn} &= - \alpha_{syn} S(V_x - \delta_{syn}), \\
		\tau_x \dot{V_x} &= V - V_x,
	\end{aligned}
\end{equation}
where $S(x) = \frac{1}{1 + \exp(-x)}$. The negative sign reflects the inhibitory nature of the synapse.

\subsection{Circuit design}

We have implemented the model outlined in the previous section on a printed circuit board, consisting of three neurons and three inhibitory synaptic connections. The implementation uses discrete MOSFET arrays for realising the membrane current elements, while parameter voltages were controlled by potentiometers. Current injection was achieved by varying the reversal potential of the membrane's leak resistance (i.e. $E_l$ in \cref{eq:membrane}).

In order to construct an HCO circuit, two of the three neurons were connected as described in section \ref{section: HCO} with mutually inhibitory synaptic connections described in \cref{eq:synapse}.

\subsection{Experimental Setup}

The setup consisted of the circuit board, a signal generator used to generate input current waveforms, and an oscilloscope that recorded the output voltages. In each experiment, different trials were recorded by leaving a time interval of 30-120 seconds between each successive voltage recording in order to observe the effects of noise and temperature variations.

For the spiking SiN experiments, it was first subjected to a constant input current just above its spiking threshold, and the voltage response was recorded for 25 different trials. Then the SiN was subjected to the same fluctuating input current in each trial, originally generated using a Gaussian white-noise process ($\mu = 1.1$ V, $\sigma = 0.02$ V) and the voltage response was recorded across 25 trials. The DC offset of 1.1 V was chosen as it is just below the spiking threshold. In the bursting SiN experiment, the same experimental procedure was carried out, even using the same fluctuating input used in the spiking SiN experiments, but now slowed down by 50 times. This adjustment reflects that the bursting period is approximately 50 times longer than the spiking period. Additionally, the DC offset of the input was adjusted so it is just below the bursting threshold.  Similarly, the input was further slowed down by 2 times in the HCO experiment, compared to the bursting experiment, as the network's anti-phase oscillations consist of two bursts per period. In each case, zero-order hold interpolation was used when adjusting the input signal for each temporal scale.

\section{Results}
\subsection{Reliable Spike Transmission}

\begin{figure}[h]
	\centering
	\begin{subfigure}{1.0\textwidth}
		\centering
		\includegraphics[width=1.0\linewidth]{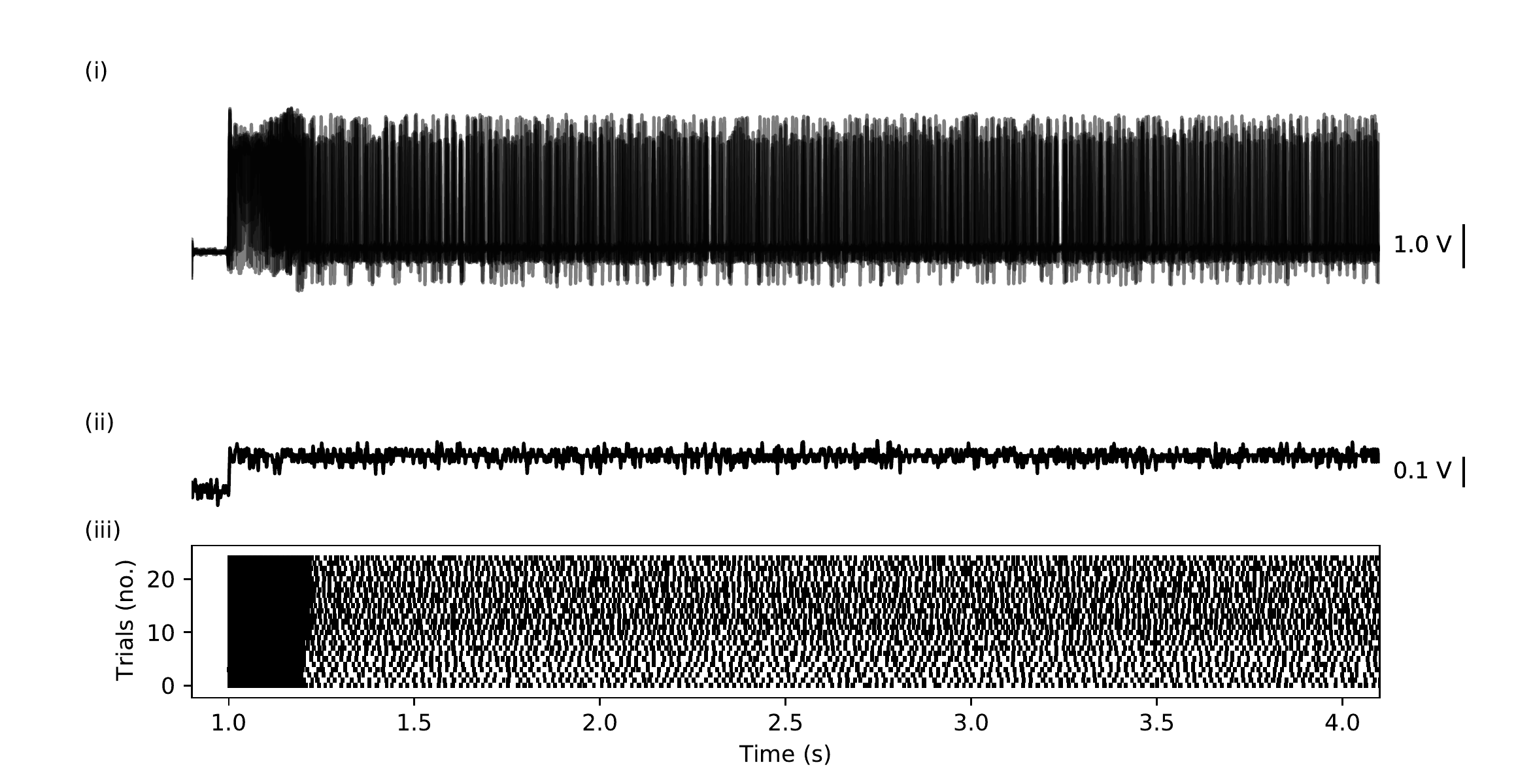}
		\caption{Step current input.}
		\vspace{0.5cm}
		\label{spike reliability:sub1}
	\end{subfigure}
	\begin{subfigure}{1.0\textwidth}
		\centering
		\includegraphics[width=1.0\linewidth]{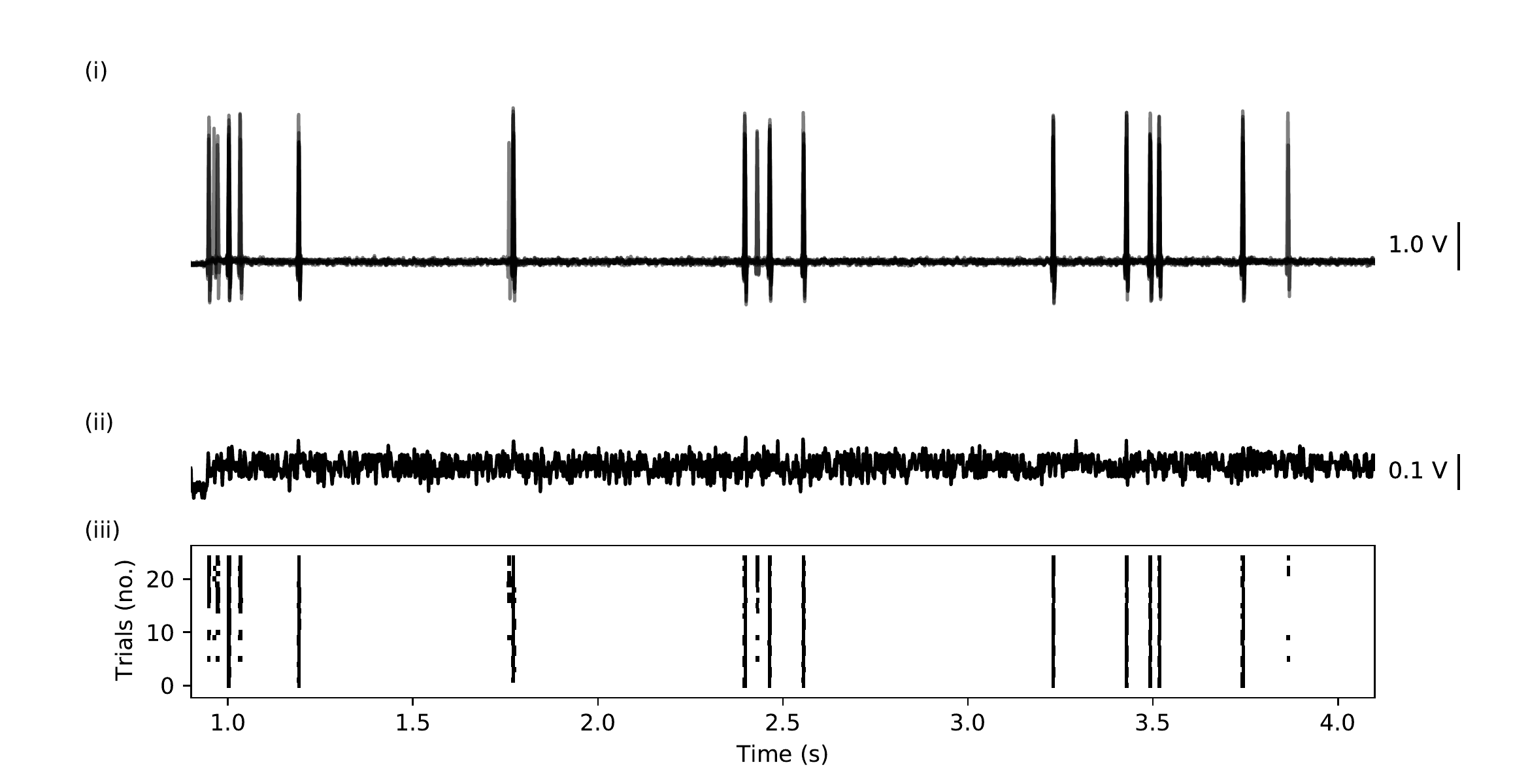}
		\caption{Fluctuating current input.}
		\label{spike reliability:sub2}
	\end{subfigure}
	\caption{\doublespacing Spike-timing reliability in the neuromorphic neuron. Each figure shows the superimposed voltage and current (shown in terms of $E_l$) signals measured in five successive trials in (i) and (ii) respectively, and a raster plot in (iii). When the neuron is subjected to a step input current (\cref{spike reliability:sub1}), the spike timings are highly inconsistent due to the noise and temperature fluctuations between trials. On the other hand, when the neuron is subjected to a fluctuating input current (\cref{spike reliability:sub2}), the output spikes appear in a reliable manner between trials.}
	\label{spike reliability}
\end{figure}

In the first experiment, we first subjected a single neuron on the board to a constant input current, leading to intrinsic spiking.  Subsequently, we injected a fluctuating current that intermittently induces output spikes. The resulting waveforms and raster plots are shown in \cref{spike reliability}.

\subsection{Reliable Burst Transmission}

\begin{figure}[h]
	\centering
	\begin{subfigure}{1.0\textwidth}
		\centering
		\includegraphics[width=1.0\linewidth]{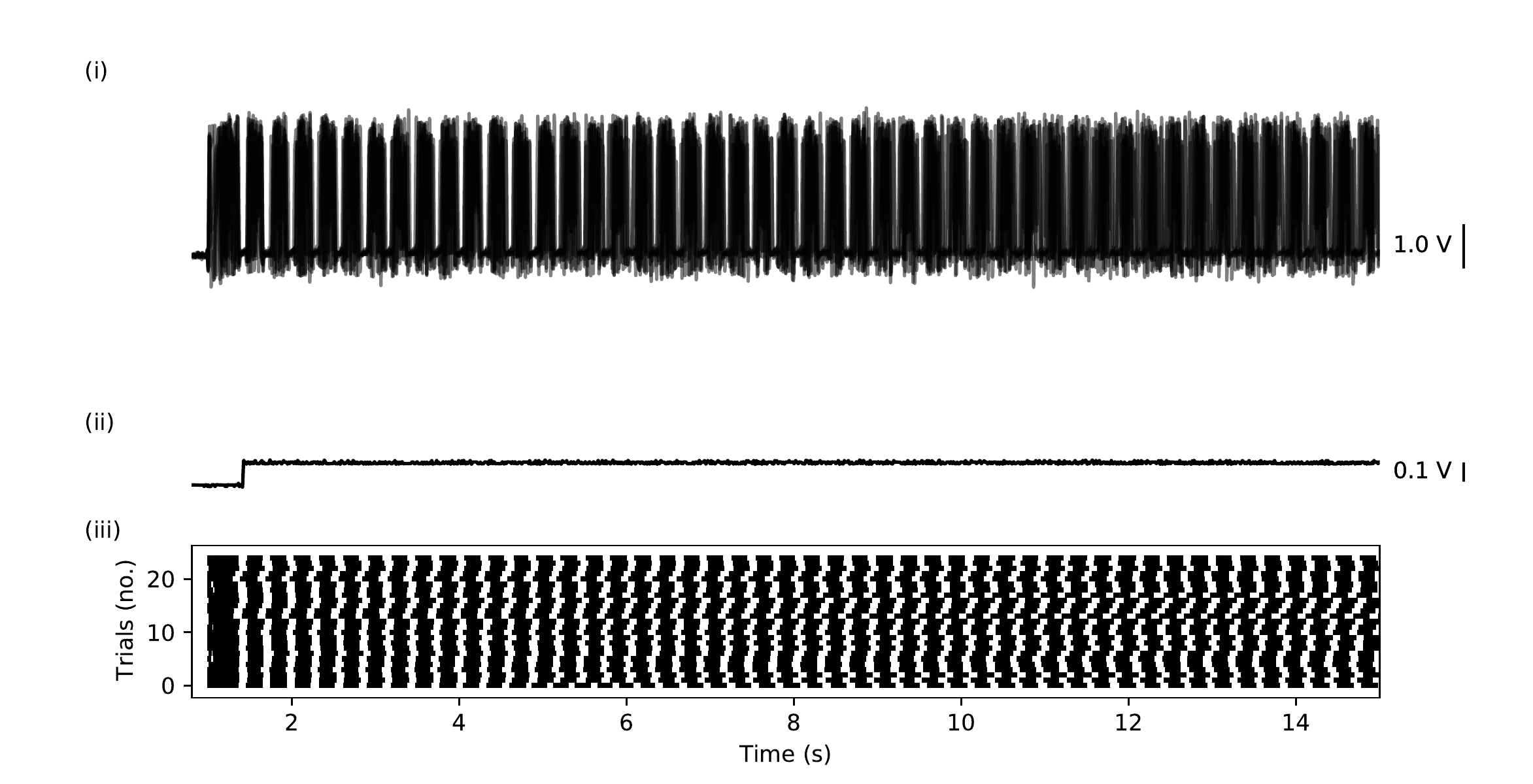}
		\caption{Step input.}
		\vspace{0.5cm}
		\label{burst reliability:sub1}
	\end{subfigure}
	\begin{subfigure}{1.0\textwidth}
		\centering
		\includegraphics[width=1.0\linewidth]{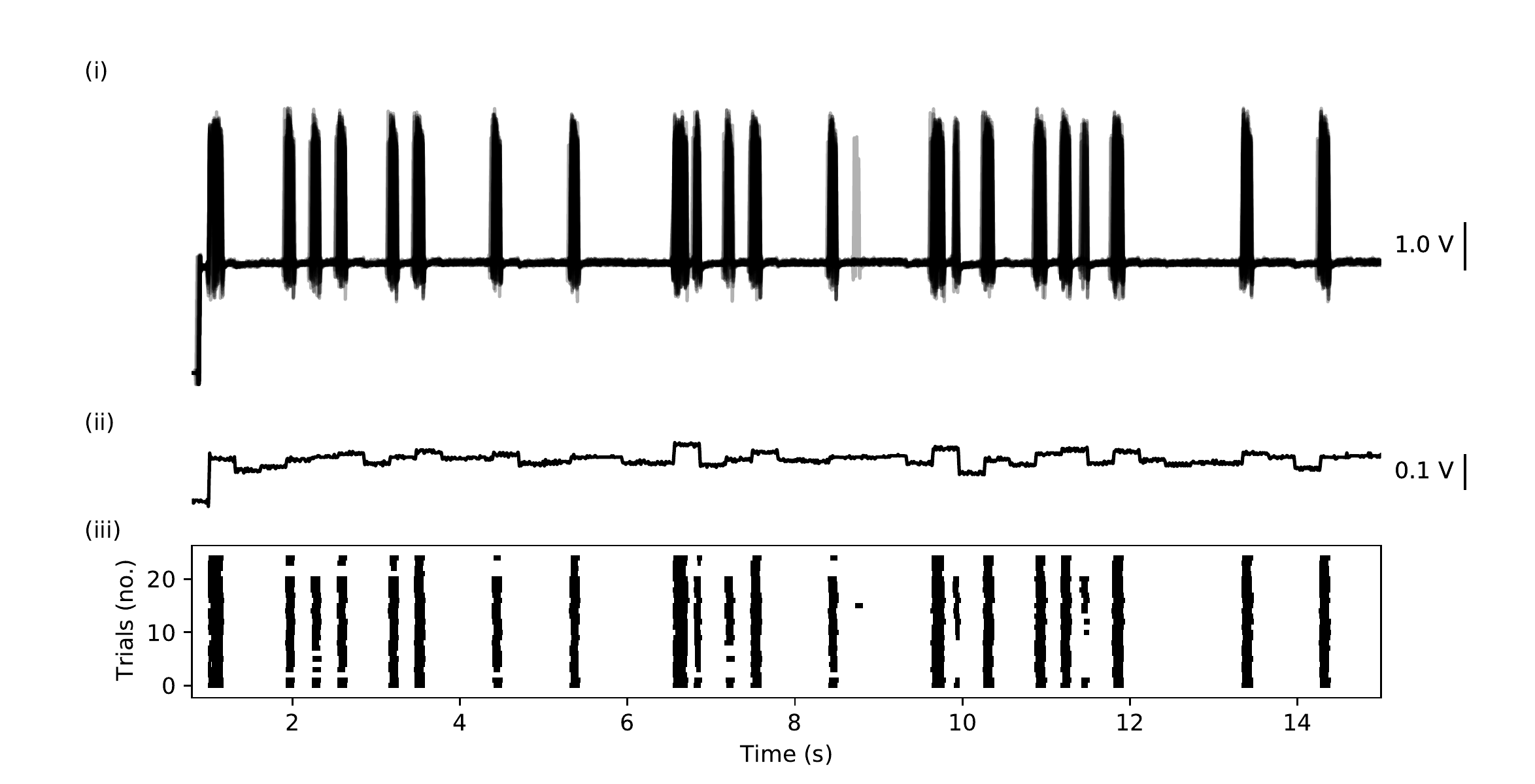}
		\caption{Fluctuating input.}
		\label{burst reliability:sub2}
	\end{subfigure}
	\caption{\doublespacing Burst-timing reliability in the neuromorphic neuron. The experiments of \cref{spike reliability} were reproduced for the neuron in bursting mode. By considering the step input current (\cref{burst reliability:sub1}), we again observe that burst timings drift over time in an inconsistent manner. However, injecting a slow fluctuating signal (\cref{burst reliability:sub2}) produces a highly consistent bursting trace between different trials. Note that since bursting is a slower phenomenon than spiking, the fluctuating input current was slowed-down accordingly to capture the correct timescale of bursting.}
	\label{burst reliability}
\end{figure}

In the second experiment, we increased the slow negative conductance gain ($\alpha_s^-$ parameter in the model) of the same neuron to switch its behaviour from spiking to bursting. The same experimental procedure was then carried out with the slowed-down version of the input signal. The resulting behaviour is shown in \cref{burst reliability}.

\subsection{Reliable Half-Centre Oscillator Control}

\begin{figure}[h]
	\centering
	\begin{subfigure}{1.0\textwidth}
		\centering
		\includegraphics[width=1.0\linewidth]{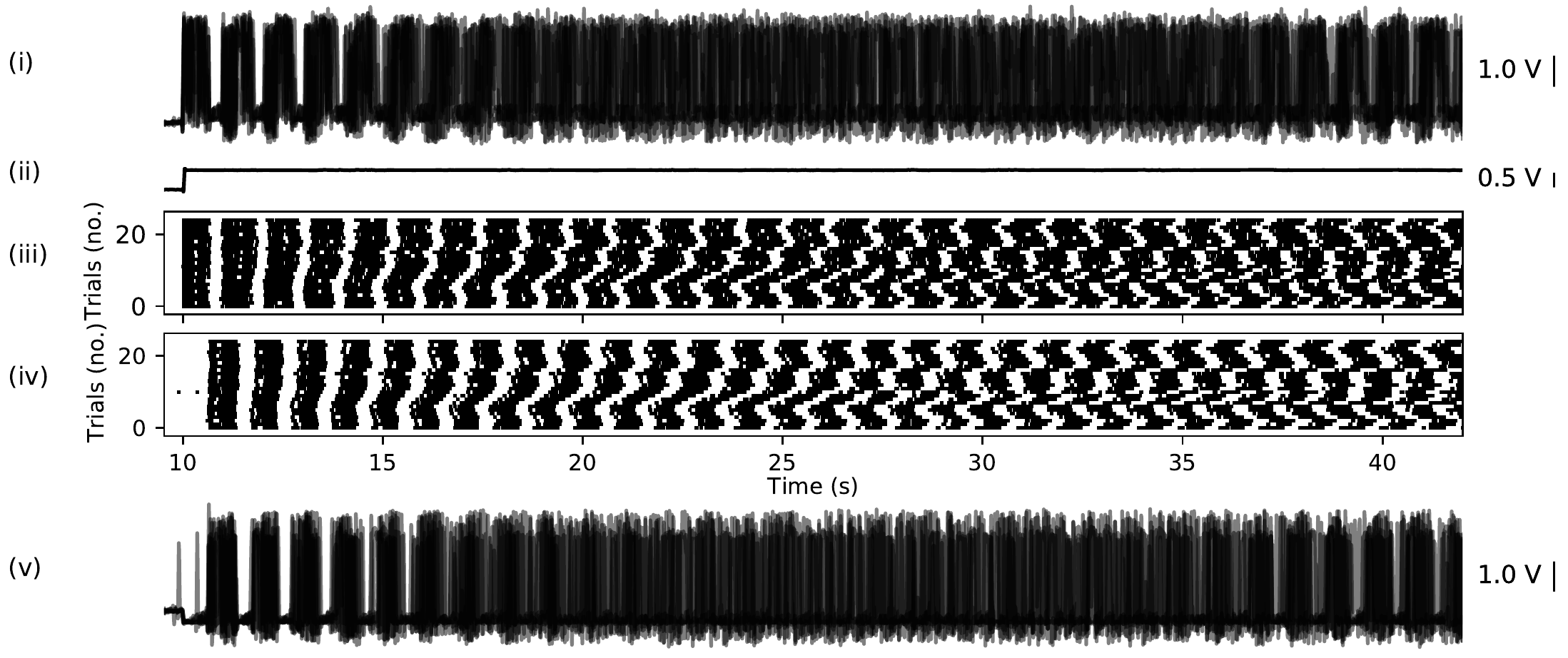}
		\caption{Step input.}
		\label{hco reliability:sub1}
		\vspace{0.5cm}
	\end{subfigure}
	\begin{subfigure}{1.0\textwidth}
		\centering
		\includegraphics[width=1.0\linewidth]{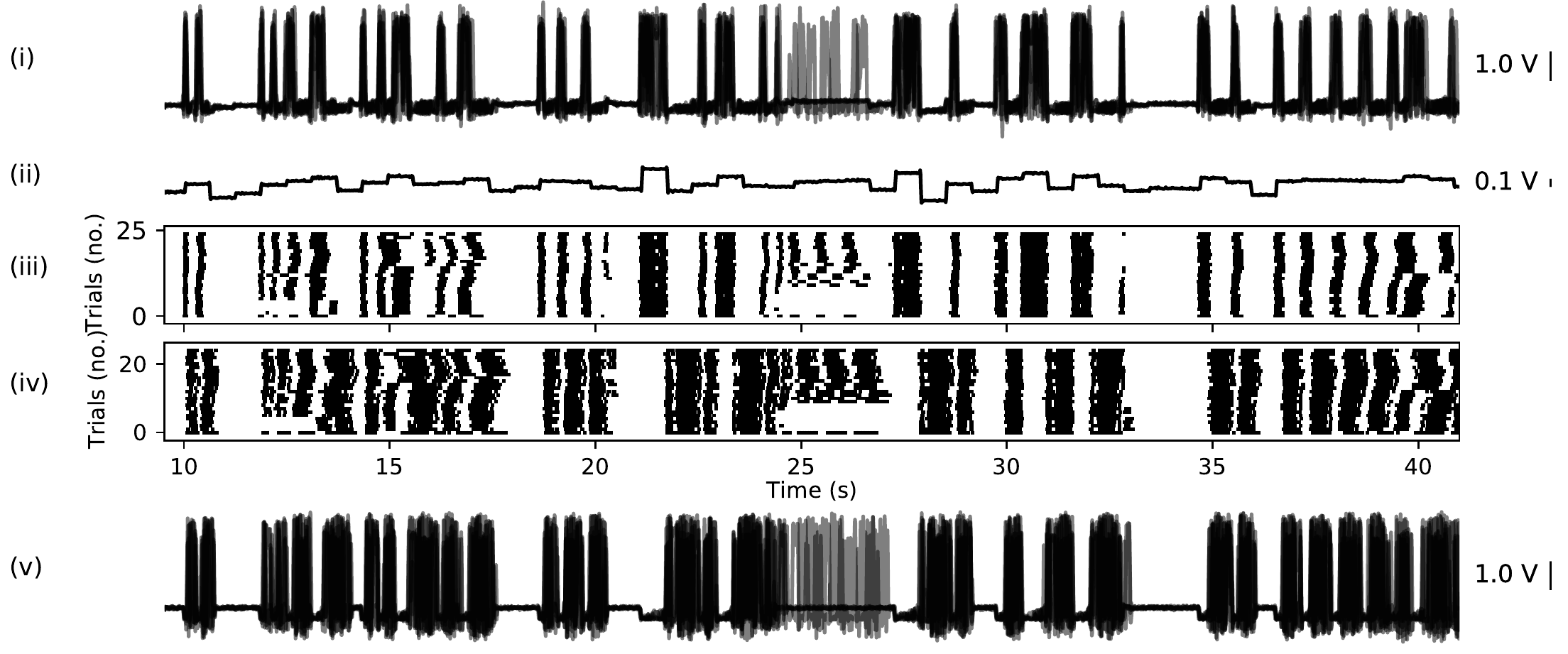}
		\caption{Fluctuating input.}
		\label{hco reliability:sub2}
	\end{subfigure}
	\caption{\doublespacing Reliability of the half-centre oscillator (HCO) circuit. In these experiments, two neurons (Neurons A and B) were connected through mutual inhibition, so that the network experiences an anti-phase oscillatory pattern. When subjecting one of the neurons (Neuron A) to an input current step (\cref{hco reliability:sub1}), the oscillations drift over time between different trials. When subjecting Neuron A to a slowly-fluctuating input (\cref{hco reliability:sub2}), the anti-phase oscillations are switched on and off in a reliable manner. In both figures, (i) and (ii) show the superimposed voltage and current over five successive trials in Neuron A, (iii) shows the raster plot for Neuron A, (iv) shows the raster plot for Neuron B, (v) shows the superimposed voltage over five successive trials in Neuron B.}
	\label{hco reliability}
\end{figure}

In the final experiment, we constructed an HCO using two neurons on the board, mutually interconnected through inhibitory synapses. A single neuron of the network was then subjected to a step change in the input current, and then the further slowed-down fluctuating current. Responses of the two neurons in the network are shown in \cref{hco reliability} (traces and raster plots).

\section{Discussion}

In the experiments carried out, we have investigated the reliability of neuronal transmission (of events) at three different temporal scales within our neuromorphic framework. Firstly, we investigated neuronal transmission at the single spike level within a single neuron (\cref{spike reliability}), mirroring the Mainen and Sejnowski experiments previously done in neocortical neurons \citep{mainen1995}. The results of those original experiments were successfully reproduced here.  In response to a step input (\cref{spike reliability:sub1}), temperature variations and noise led to a clear variability in spike timings between trials, after the initial adaptation period. In contrast, a fluctuating input signal (\cref{spike reliability:sub2}) led to highly consistent spike timings over successive trials, suggesting reliability of spike timings when each spike is a supra-threshold event response to the input.

Similarly, we investigated the reliability/robustness of the bursting SiNs (\cref{burst reliability}). In this case, the individual events are \textit{bursts} that are generated on a significantly slower timescale than the individual spikes. The reliability of burst transmission was analysed in exactly the same way by considering the neuron's response over a longer duration, to a slowed-down version of the input current. Comparing, again, the response to a constant stimulus (\cref{burst reliability:sub1}) and a slow fluctuating input (\cref{burst reliability:sub2}), we obtain similar results - the response of the neuron to a fluctuating input is highly reliable (most bursts appear at the same times across different trials). Notice that the response to a constant stimulus shows less variability than in the spiking case as bursts are generated on a significantly slower timescale and thus, it takes a longer time for them to appear out of sync.  Additionally, the mechanism for intrinsic bursting appears to be more robust than that for intrinsic spiking.

Finally, we investigated reliability at the network scale, by considering the response of the HCO to a current injected into a single neuron in the network (\cref{hco reliability}). At the HCO level, the events are anti-phase oscillations, each oscillation consisting of two bursts (one from each neuron). The anti-phase oscillations can be controlled by modulating the input current to any of the neurons in the network.  The single neuron control was able to reliably trigger network events (\cref{hco reliability:sub2}), similarly to the single spiking and bursting SiNs, but with a slower fluctuating input. Comparatively, the intrinsic oscillations (triggered by the DC current pulses) drift over time (\cref{hco reliability:sub1}). These results suggest that the behaviour of the network can be reliably controlled at a single neuron level even when variability/noise is present, indicating a multi-scale control principle that can be generalized to different network structures.

Our results suggest that analog neuromorphic circuits can be made reliable similarly to their biological counterparts. They additionally suggest that reliability can hold at different temporal scales.  This reliability property crucially depends on the robustness of an excitable system's \textit{threshold} to temperature variations and noise. This robustness enables the system to generate output events in response to excitatory inputs in a consistent manner. Although our neuromorphic implementation suffered from imperfections inherent to any analog computational devices, the existence of a well-defined and robust \textit{fast threshold} (voltage level above which individual spikes are generated), and of a \textit{slow threshold} (voltage level above which bursts are generated), allowed for a reliable input-output behaviour, even in the network setting \citep{ribar2019}. In that sense, robust thresholds are what determine the reliability of excitable systems, whether in vivo or in silico. In earlier work, we have shown that robust thresholds arise from negative conductance elements, and that models lacking those elements usually do not have robust thresholds \citep{franci2017}.

\section{Conclusion}

In this letter we have investigated the spike transmission reliability properties of an analog neuromorphic circuit at different temporal scales. By studying reliability at the level of individual spikes, bursts, and network oscillations, we have shown that the concept of reliability is an input-output characteristic of any event-based signalling system, that can hold at different temporal scales. Our results indicate that, similarly to how biological neurons exhibit reliable spiking despite their inherent variability and noise, analog neuromorphic implementations can exhibit the same reliability properties despite their hardware imperfections. This property suggests that designing reliable neuromorphic hardware out of variable and imperfect analog components is possible. Like their biological counterparts, neuromorphic designs could therefore make analog variability a feature, rather than a design hurdle, as long as event-based reliability is also established.

\subsection*{Acknowledgements}
This work was supported by the European Research Council (ERC) under the Advanced ERC Grant Agreement Switchlet n.670645.


\clearpage
\bibliography{references}
\bibliographystyle{apalike}

\end{document}